\documentclass[reprint, amssymb, nobibnotes, aip, jcp, 10pt, superscriptaddress]{revtex4-1}
\usepackage{graphicx}
\usepackage{tabularx}
\newcolumntype{Y}{>{\raggedleft\arraybackslash}X}
\usepackage[T1]{fontenc}
\usepackage[utf8]{inputenc}
\usepackage{amsmath}
\usepackage{cleveref}
\usepackage{siunitx}

\begin{document}
\title{Free energy landscape of dissociative adsorption of methane on ideal and defected graphene from ab initio simulations}
\author{M. Wlaz{\l}o}
\email{mateusz.wlazlo@fuw.edu.pl}

\author{J.A. Majewski}
\affiliation{Faculty of Physics, University of Warsaw, Pasteura 5, 02-093 Warsaw, Poland}
\date{\today}

\begin{abstract}
We study the dissociative adsorption of methane at the surface of graphene. Free energy profiles, which include activation energies for different steps of the reaction, are computed from constrained \textit{ab initio} molecular dynamics. At 300~\si{\kelvin}, the reaction barriers are much lower than experimental bond dissociation energies of gaseous methane, strongly indicating that graphene surface acts as a catalyst of methane decomposition. On the other hand, the barriers are still much  higher than on nickel surface. Methane dissociation therefore occurs at a higher rate on nickel than on graphene. This reaction is a prerequisite for graphene growth from precursor gas. Thus, the growth of the first monolayer should be a fast and efficient process while subsequent layers grow at diminished rate and in a more controllable manner. Defects may also influence reaction energetics. This is evident from our results, in which simple defects (Stone-Wales defect and nitrogen substitution) lead to different free energy landscapes at both dissociation and adsorption steps of the process.
\end{abstract}


\maketitle 

\section{Introduction}
The process of activated adsorption of methane on transition metals has been under thorough experimental and theoretical examination. Pioneering work in this field was reported by Winters in 1975 \cite{Winters1975}. Results obtained from kinetic studies of methane adsorption on atomically clean tungsten surface placed in ultrahigh vacuum chamber showed that the dissociative chemisorption reaction is favored:
\begin{equation*}
	CH_4(gas) \rightarrow CH_x(ads)+(4-x)H(ads).
\end{equation*}

A number of different transition metal surfaces was studied in a similar manner, including nickel (100), (110) and (111) faces \cite{CHn_kin_Ni1,CHn_kin_Ni2,CHn_kin_Ni3}. All these studies indicated a high temperature dependence of dissociation probabilities of CH$_4$ on Ni. More intricate details of the reaction mechanism on Ni have been unraveled in molecular beam experiments \cite{CHn_molbeam_Ni1,CHn_molbeam_Ni2,CHn_molbeam_Ni3,CHn_molbeam_Ni4} and in state-resolved measurements \cite{CHn_state_Ni1,CHn_state_Ni2} that attributed increased reactivity to excitations of specific modes in CH$_4$ ($\nu_3$ and $2\nu_3$). Also the sticking probabilities obtained from first-principles calculations in the framework of density functional theory (DFT) calculations were in fairly good agreement with experimental data \cite{CHn_DFT_Ni1,CHn_DFT_Ni2}. A detailed review of the advancements in the subfield of dissociative chemisorption dynamics on metals can be found in Ref.~\onlinecite{CHn_review}.

CH$_4$ decomposition reactions on other metals were also studied with DFT in a few earlier papers, namely on copper \cite{Cu1}, iridium \cite{Ir}, and copper alloys, Cu-Fe \cite{Fe-Cu} and Ni-Cu \cite{Ni-Cu}. 

DFT calculations of adsorption (in fact physisorption) of CH$_4$ on graphene have also been carried out \cite{ch4-1, ch4-2, ch4-3}. The CH$_4$ decomposition reactions on graphene were recently studied in the framework of classical molecular dynamics \cite{clasMD}. To this day, however, we are not aware of such studies in the framework of AIMD.

In recent years, catalytic dissociation of CH$_4$ on metals has been utilized in graphene CVD growth process. In particular, nickel is a suitable substrate for growth of large, high-quality graphene layers \cite{Graphene_CVD_Ni}. It allows for fabrication of monolayer and few-layers-thick graphene. Controlled bilayer graphene growth on Cu-Ni alloys has also been demonstrated \cite{Bilayer_CVD_Ni}. Theoretically, dissociation of methane on Ni(111) surface has been studied with \textit{ab initio} molecular dynamics (AIMD) and activation energies for dissociation steps have been computed \cite{PMF_CHn-diss-Ni}. These studies described just the initial step of the growth process that has to occur before the first layer can be formed.

In this study we use an \textit{ab initio} framework to compute accurate and realistic free energy profiles for a series of subsequent steps which amount to dissociative chemi- and physisoprption of methane on monolayer graphene. The study profits from high performance computing capabilities. They are utilized to provide long simulation times and system size that allows to capture interesting surface phenomena. In particular, we study two types of chemical processes. We consider first the sequential dehydrogenation of methane and then the adsorption process of methane fragments CH$_n$ ($n = 1,2,3$) to the graphene layer. We consider the above-mentioned processes in pristine and defected graphene and point out that the structural defects, such as the Stone-Wales defect or nitrogen substitutional impurity, influence the dynamics of methane's fragments dehydrogenation and adsorption. In the context of growth processes, the considered chemical reactions are crucial for the growth of the second graphene layer in a chemical vapor deposition process with methane as precursor.

In previous studies, graphene geometry on Ni(111) substrate has been studied experimentally by low-energy electron diffraction and by modelling in the DFT framework. While the interlayer separation is around 2.0-2.1~\si{\angstrom}, there is no evidence of $sp^2$ to $sp^3$ rehybridization \cite{GrapheneMetal}. The adsorption energy of a graphene monolayer on nickel is typical for physisorption \cite{GrapheneNiAdsorption}. For a similarly bound graphite system, it has been shown that neighboring carbon layers have weak influence on adsorption energy \cite{GraphiteInfluence}. It is therefore reasonable to assume that a model that contains only the outermost atomic layer is sufficient to capture adsorption phenomena that occur on nickel-suspended graphene with sufficient accuracy.

The determination of energy barriers for CH$_4$ decomposition in subsequent dehydrogenation steps and comparison with the corresponding barriers for CH$_4$ decomposition on Ni surface (taken from Ref.~\onlinecite{PMF_CHn-diss-Ni}) will allow us to compare the rates at which C atoms are delivered for graphene growth at each surface. Faster CH$_4$ decomposition on Ni should favor the graphene monolayer growth, and the faster decomposition on graphene should favor growth of islands. Since it is already known that monolayers of graphene are obtained in the CVD process on Ni surface, we expect that the energy barriers for the CH$_4$ decomposition are lower for the nickel surface than for graphene. Nevertheless, confirmation of this fact on the basis of quantum-mechanical theory and determination of the catalytic features of graphene for important chemical reaction such as CH$_4$ decomposition seems to be of interest for general field of chemical reactions of gas-phase molecules on solid surfaces. 

The calculation scheme we employ in simulations is detailed in \Cref{section:Methods_description}, where the geometry of the calculated systems is also described. Further on in \Cref{section:Results}, we demonstrate and discuss obtained free energy profiles of decomposition (\Cref{section:decomposition}) and adsorption (\Cref{section:adsorption}) of methane and related species on graphene. We also examine electron density distribution maps (\Cref{section:density}) of pristine graphene and graphene with admolecules, which provide us some hints how the surface reacts to adsorption of different species.

\section{Methods}
	\label{section:Methods_description}

	\subsection{Computational approach}
	Reaction free energies have been computed from AIMD simulations. To sample the reactions with high precision but long computation times, a scheme to accelerate sampling has been used. There are several ways to address the issue of slow sampling in AIMD, each with their strengths and drawbacks \cite{chipot,PMF_methods}. For the system in question, we have selected the constraint-based simulation that ensures uniform sampling of the relatively simple reaction coordinate (RC).

	The graphene monolayer has been simulated with a 4x4 hexagonal supercell with unit cell parameters $a=9.84~\si{\angstrom}$ and $c=20~\si{\angstrom}$. Cell parameter $a$ corresponds to most often reported experimental lattice constant of graphene (2.46~\si{\angstrom} for the primitive unit cell). High value of $c$ ensures minimal interaction between neighboring periodic images. The resulting layer of vacuum is wider than typically found in literature for similar systems \cite{GrapheneNiAdsorption,Fe-Cu,Ni-Cu}. The system is comprised of 32 in-plane carbon atoms and a single adsorbate molecule (left side of reaction equations). The full system geometry is visible in \Cref{figure:ch3_diss}. The size of the supercell allows for relaxation of the system towards the most stable atomic configuration, therefore encapsulating the most important effects. As CH$_n$ molecules are chemisorbed to the surface, the hybridization of the in-plane carbon changes from $sp^2$ to $sp^3$. With that change, the local geometry shifts from planar to buckled. This affects also the neighboring atoms that are pulled slightly above the graphene plane. The present choice of the supercell guarantees that this effect is accounted for properly. A smaller supercell would hinder this movement resulting in strained geometry of the simulated system.

	AIMD simulations have been performed with holonomic constraints imposed on the RC. A constraint force is introduced at each time step to keep the RC fixed. At the end of each simulation, the time average of the force is computed. Relative free energies of the reactions are calculated as the potential of the mean force (PMF) by numerical integration. The theoretical basis for this scheme had been laid by Carter et al. \cite{PMF_Carter} and then it received a number of developments \cite{PMF_Mulders,PMF_Otter,PMF_Sprik,PMF_Otter2,PMF_Pohorille,PMF_Schlitter}. This approach is rooted in the thermodynamic integration approach first conceived by Kirkwood in 1935 \cite{Kirkwood}. Integration is done from low to high RC values.
	
	The constraint is introduced as an additional term in the extended Lagrangian of the system:

	\begin{equation*}
		\mathcal{L}^*(\mathbf{q},\mathbf{\dot{q}}) = \mathcal{L}(\mathbf{q},\mathbf{\dot{q}}) + \sum_{i=1}^{N_c}\lambda_i\sigma_i(\mathbf{q}).
	\end{equation*}

	In the last term, $\sigma_i$ defines the geometric constraints, $N_c$ is the number of constraints employed, and $\lambda_i$ are the associated Lagrange multipliers. These can be interpreted as the force that is required to keep the geometric constraint fixed. They are recalculated at each time step using the SHAKE/RATTLE \cite{SHAKE,RATTLE} algorithm. In principle, an arbitrary number of constraints can be included in the Lagrangian, but for the purpose of the present study, we limit ourselves to just one constraint.
	
	Over the course of the simulations, data on $\lambda_i$ is gathered and the mean force (MF) is computed. To obtain the free energy profile, we calculate the potential of mean force (PMF) by integrating the MF:

	\begin{equation*}
		PMF=-\int_{r_1}^{r_2}\left<\lambda(r)\right>dr
	\end{equation*}

	Integration bounds r$_1$ and r$_2$ are chosen arbitrarily for each reaction. The upper bound r$_2$ is in the range of 3.1-4.0~\si{\angstrom}. Depending on the species, the free energy profiles show different asymptotic behavior at high RC values. In some cases, $\lambda_i$ vanishes exactly at high RC, which produces a flat tail in the free energy profile. If the opposite is true, an ascending or descending tail appears. This suggests spurious interaction that would vanish given a high enough r$_2$ value.

	Positive values of the constraint force mean that the constraint holds the molecule together, when it would otherwise decompose. Negative values mean that the constraint pushes the atoms apart. The equilibrium value of the constraint corresponds to the free energy minimum, where the force vanishes and changes sign from positive to negative.

	In our studies the calculations are performed as follows. First, the graphene backbone is equilibrated for 1.8~\si{\pico\second} without adsorbate molecules. Then, the calculation is split into segments for each value of the reaction coordinate, and the following procedure is carried out for each segment:

	\begin{itemize}
		\item Initial positions of the in-plane carbon atoms are taken from the equilibration run. The adsorbate molecule is then added close to the surface and the ground state is found using DFT.
		\item The system is equilibrated for 1500 MD steps, and then the simulation is started with ground state electronic density found in step 1 and velocities from graphene backbone equilibration.
		\item After equilibration, ground state calculation is performed once more.
		\item The MD simulation is started from the new ground state and equilibrated positions and velocities. The production takes 15000 additional MD steps.
	\end{itemize}

	At the time step of 4 atomic units (0.096~\si{\femto\second}), the equilibration period is equal to 144~\si{\femto\second} and the production run during which data is gathered takes 1.44~\si{\pico\second}. Total simulation time is equal to 43.20~\si{\pico\second} for each reaction step.

	For molecular dynamics runs, we use the Car-Parrinello method \cite{Car-Parrinello}. All equilibrations are performed in the canonical ensemble using the Nos\'e-Hoover chain thermostat \cite{nosethermostat,hooverthermostat}. Ionic thermostat is set to target temperature of 300~\si{\kelvin}.

	Forces required to perform the MD simulations are Hellmann-Feynmann forces obtained via plane wave DFT \cite{DFT-HK,DFT-KS} calculations carried out with the CPMD package \cite{CPMD}. Gradient-corrected Becke-Lee-Yang-Parr \cite{Becke,LYP} approximation of the exchange-correlation functional is used along with Troullier-Martins-type norm-conserving pseudopotentials \cite{MT_PP} constructed by Boero for hydrogen and carbon \cite{Boero_H_BLYP}. Kohn-Sham orbitals are expanded in the plane-wave basis set with the kinetic energy cutoff of 70~Ry, on par with previous studies of hydrocarbon systems \cite{Cutoff_Draxl,Cutoff_Galli}. The DFT-D2 semi-empirical dispersion correction by Grimme \cite{DFT-D2} is used to account for van der Waals interaction. Brillouin Zone sampling was limited to the $\Gamma$ point. See Supplemental Information for the results of convergence testing with respect to k-point grid density. Tests reveal fair convergence but as integration errors accumulate along the path the behavior at high RC values becomes volatile.

\section{Results and discussion}
\label{section:Results}
Before outlining our findings, let us briefly describe our system geometries. In our calculations we have included pristine graphene (PG) geometry and two defected geometries~--~Stone-Wales-defected graphene (SWG) and N-substituted graphene (NSG).
\begin{itemize}
	\item SWG is obtained from PG by a simple rearrangement of atoms which involves rotating one of the C-C bonds by 90 degrees \cite{SW-CPL}. This introduces strain that is released by out-of-plane buckling around the defect \cite{SW-DFT-QMC-buck}. In a previous study of adsorption energetics via static DFT calculations we have investigated adsorption profiles of $H$ and CH$_3$ on different atomic sites of SWG \cite{APPA}. Adsorption sites close to the SW defect were favored over more distant ones. In this work we attempt to see if earlier conclusions extend to more realistic scenarios simulated here.
	\item NSG features a single substitution of a carbon atom in graphene with a nitrogen atom. N-doped graphene has been demonstrated to have exceptional catalytic properties towards various reduction reactions \cite{N_defect}. Here we aim to see if this effect translates to a simulation of reactions occurring in the vicinity of a single in-plane dopant atom.
\end{itemize}

Obtained relative free energy profiles are depicted in \Cref{figure:pmf_diss,figure:PMF_ads}. In addition, \Cref{table:diss,table:ads} list RC values and energy barriers for quick comparison. Each energy profile is shifted so that zero energy corresponds either to the global energy minimum (in case of decomposition reactions), or the chemisorption minimum (for adsorption reactions).

	\subsection{CH$_n$ decomposition}
	\label{section:decomposition}
	\begin{figure}
	\centering
	\includegraphics[width=3.37in]{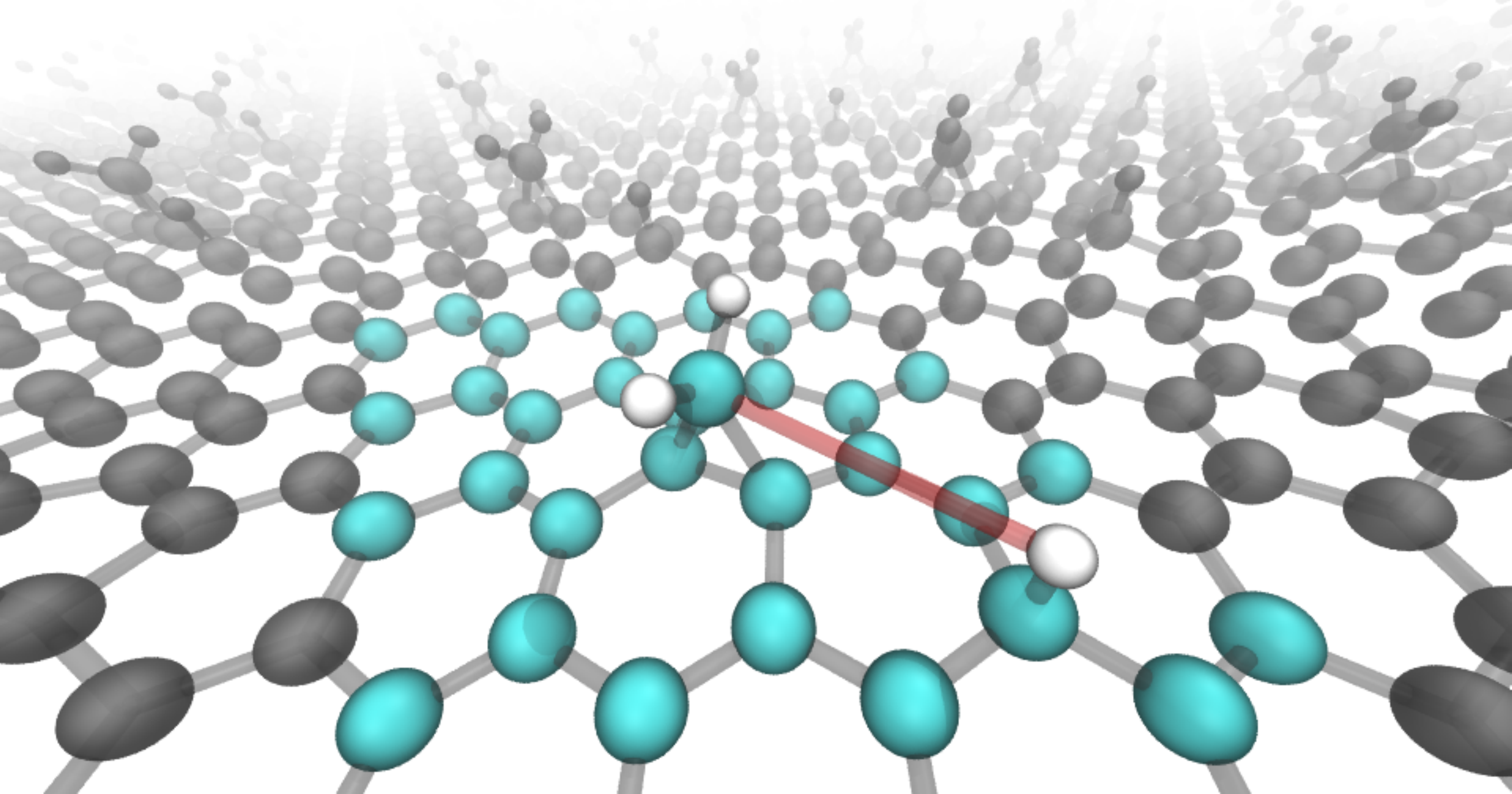}
	\caption{CH$_3$ dissociated into CH$_2$ and H on graphene at RC equal to 3.62~\si{\angstrom} (red transparent bond). This is the final atomic configuration from the production run. The free energy profile for this reaction is plotted in \Cref{figure:pmf_diss}(a) and labeled `$CH_3 \rightarrow CH_2+H$'. The magnitudes of energy barriers and corresponding RC values are listed in \Cref{table:diss}. Color description: Colored atoms are included in the supercell (teal~--~carbon, white~--~hydrogen), and surrounding gray atoms are periodic images of the supercell. See Supplemental Information for graphics depicting configurations of other reaction steps.}
	\label{figure:ch3_diss}
	\end{figure}
	We start presentation of results with catalytic methane decomposition on graphene. The process, starting from the methane molecule (CH$_4$) and ending with the C atom, can be broken down into four reaction steps:

	\begin{align}
		CH_4 &\rightarrow CH_3+H \\
		CH_3 &\rightarrow CH_2+H \\
		CH_2 &\rightarrow CH+H \\
		CH   &\rightarrow C+H
	\end{align}

	As the product of each step, we obtain a $CH_{n-1}$ and H radicals chemisorbed at the surface (as seen e.g. in \Cref{figure:ch3_diss} and \Cref{figure:ch3_ads}). Each simulation has been started with the CH$_n$ molecule close to the surface. For CH$_4$, the starting position was chosen arbitrarily. For other molecules, the final configuration of the previous reaction step was taken as the initial position. The bond length between one of the hydrogen atoms and the carbon was selected as the RC. The rest of the C-H and C-C bonds in the system are allowed to fluctuate freely.

	\begin{table*}
		\caption{Free energy barriers and reaction coordinates (RCs) for decomposition of methane gas constituent species. Values listed here have been obtained from the data presented in \Cref{figure:pmf_diss} by third order interpolation. Experimental bond dissociation energies (BDE) for isolated species in gaseous state are from Ref.~\onlinecite{BDE1994,BDE1999}. Results of ab initio molecular dynamics (MD) and nudged elastic band (NEB) are from Ref.~\onlinecite{PMF_CHn-diss-Ni}. `--' in the last column indicates that at the upper integration bound the energy profile is ascending or descending and no second minimum is found.}
		\centering
		\begin{tabularx}{\linewidth}{l|rYY|YYYY}
			\hline
			Species & BDE (\si{\electronvolt}) \cite{BDE1994,BDE1999} & \multicolumn{2}{c|}{\vtop{\hbox{\strut Decomposition barrier}\hbox{\strut on Ni(111) (\si{\electronvolt}) \cite{PMF_CHn-diss-Ni}}}} & Decomposition barrier (\si{\electronvolt}) & Transition state RC (\si{\angstrom}) & Activation energy for reverse reaction (\si{\electronvolt}) & Second minimum RC (\si{\angstrom})\\
			& & NEB (0~\si{\kelvin}) & MD (1500~\si{\kelvin}) & & & &\\
			\hline
			$CH_4@PG$   & 4.55 & 0.78 & 0.72 & 1.37 & 1.94 & 0.50   & 3.04  \\
			$CH_4@SWG$  &      &      &      & 0.84 & 1.67 & 0.68   & 3.26  \\
			$CH_4@NSG$  &      &      &      & 1.81 & 2.28 & 0.00   &   --  \\
			$CH_3@PG$   & 4.79 & 0.81 & 0.47 & 2.03 & 2.52 &   --   &   --  \\
			$CH_3@SWG$  &      &      &      & 2.07 & 2.47 &   --   &   --  \\
			$CH_3@NSG$  &      &      &      & 2.03 & 2.36 & 0.05   & 2.81  \\
			$CH_2@PG$   & 4.39 & 0.27 & 0.23 & 2.04 & 2.38 & 0.00   &   --  \\
			$CH_2@SWG$  &      &      &      & 1.33 & 1.97 &   --   &   --  \\
			$CH_2@NSG$  &      &      &      & 1.97 & 2.87 &   --  	&   --  \\
			$CH@PG$     & 3.51 & 1.23 & 0.42 & 1.68 & 2.17 & 0.07   & 2.63  \\
			$CH@SWG$    &      &      &      & 1.37 & 1.92 & 0.42   & 2.75  \\
			$CH@NSG$    &      &      &      & 2.03 & 2.51 & 0.00   &   --  \\
			\hline
		\end{tabularx}
		\label{table:diss}
	\end{table*}

		\subsubsection{Pristine graphene}
		\begin{figure}
		\centering
		\includegraphics[width=3.37in]{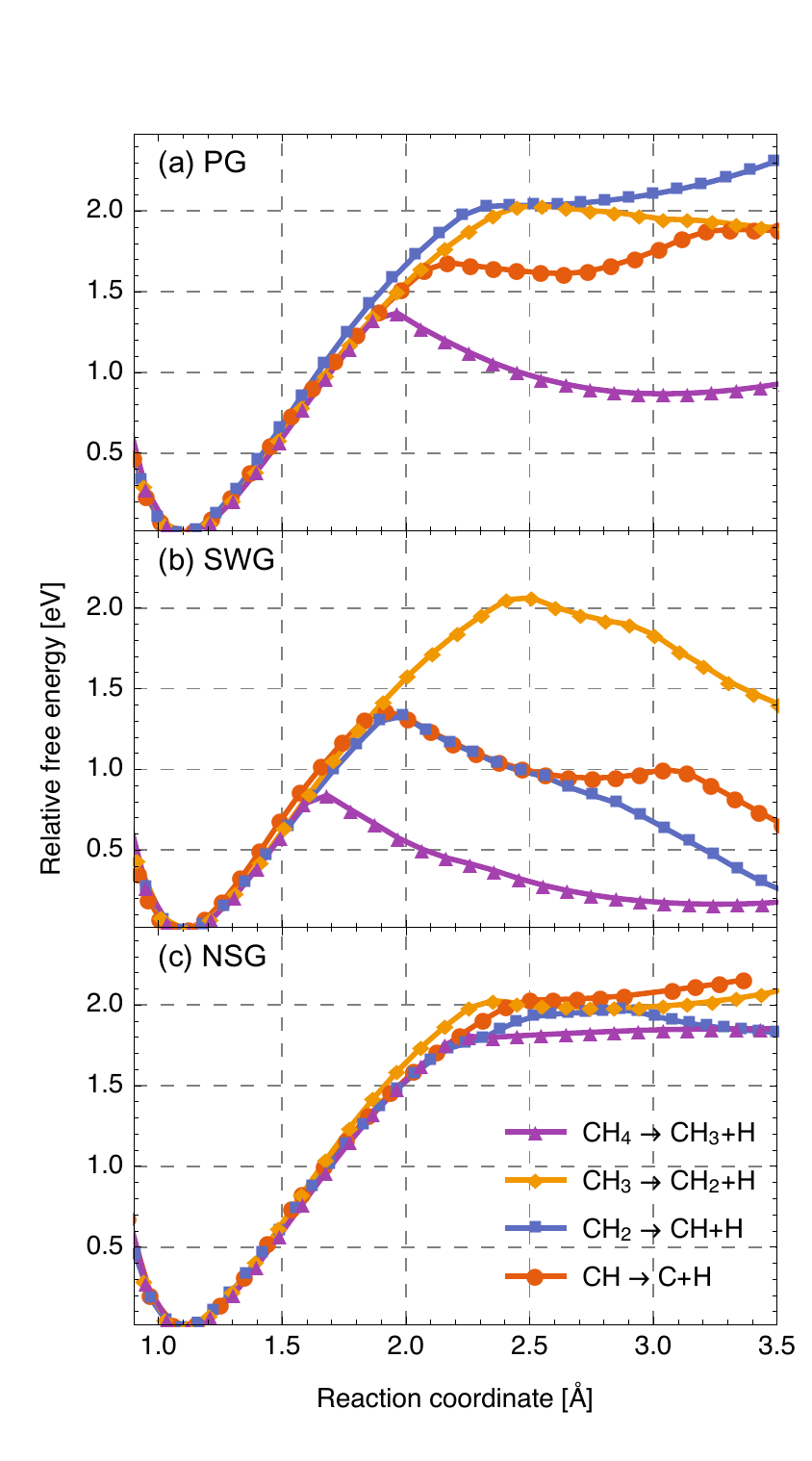}
		\caption{Relative free energy profiles of CH$_4$ dissociation steps on (a) pristine graphene, (b) graphene with a SW defect, (c) graphene with a N substitutional impurity. Curves have been shifted so that zero energy corresponds to global energy minima. The C-H bond distance is the reaction coordinate, as depicted in \Cref{figure:ch3_diss} for the $CH_3 \rightarrow CH_2+H$ reaction.}
		\label{figure:pmf_diss}
		\end{figure}

		In the first AIMD simulation series, we have studied CH$_4$ decomposition on PG. The reaction barriers for subsequent steps are listed in \Cref{table:diss}. They range from 1.37~\si{\electronvolt} for CH$_4$ up to ~2~\si{\electronvolt} for CH$_2$ and CH$_3$. Experimental bond-dissociation energies of the C-H bonds in gaseous species range between 3.51 and 4.79~\si{\electronvolt} \cite{BDE}. This means that the graphene surface has a significant catalytic effect in hydrocarbon dissociation reactions.

		 The free energy profiles for subsequent steps of CH$_4$ decomposition on graphene along the reaction coordinate (as seen for CH$_3$ decomposition in \Cref{figure:ch3_diss}) are depicted in \Cref{figure:pmf_diss}(a). They all feature a deep minimum at the equilibrium C-H bond length of 1.1~\si{\angstrom}. The initial slope of the free energy profile upon leaving the equilibrium is very similar for every step. The magnitude of the energy barriers, the presence and depth of a second free energy minimum depends on the reaction in question, as described below.

		\begin{enumerate}
			\item The first reaction is the removal of hydrogen atom from the complete methane molecule (CH$_4$). The energy barrier for removal is the lowest among the four reactions (Eqs.~(1)-(4)) and equal to 1.37~\si{\electronvolt}. The transition state is located at RC = 1.94~\si{\angstrom}. The considerable depth of the second energy minimum shows that the reaction leads to a stable product. The reverse reaction for this step is the least probable.
			\item The stability of the product of reaction (1) is also evident in the energy profile of the second decomposition step. Together with CH$_2$ decomposition barrier, it is the highest among the four reaction stages and equal to 2.04~\si{\electronvolt}. The transition state is located at 2.52~\si{\angstrom}. In the simulated RC range, the system does not reach a second free energy minimum.
			\item The energy barrier for H removal from CH$_2$ is similar to the case of CH$_3$ with an even higher transition state at 2.38~\si{\angstrom}. The product is highly unstable, as evident by the lack of second energy minimum. The product can easily undergo a reverse reaction. This likely makes CH$_2$+2H the most stable decomposed state of methane on graphene.
			\item The last removal of H has a slightly lower energy barrier of 1.68~\si{\electronvolt}. The transition state is only slightly higher than for CH$_4$ at 2.18~\si{\angstrom}. The second energy minimum is deeper than for CH$_2$. Thus the reverse reaction should be less probable than for the previous step.
		\end{enumerate}
		
		\subsubsection{Stone-Wales-defected graphene}
		In the subsequent series of calculations, reactions (1)-(4) have been simulated for the case of graphene with the Stone-Wales defect (i.e., with the supercell containing the SW defect). Initially, the CH$_4$ molecule has been placed directly over the defect site. The resulting free energy profiles depicted in \Cref{figure:pmf_diss}(b) are quite different in both shape and magnitude of energy barriers as compared to the case of dehydrogenation reactions (Eqs.~(1)-(4)) at pristine graphene that have been discussed above. 

		The first stage reaction (Eq.~(1)) has an even lower energy barrier compared to the same reaction occurring on pristine graphene at 0.84~\si{\electronvolt}. In fact, this barrier is the lowest among all studied reactions for all investigated configurations. The barrier is also lowered by a slightly lesser degree in case of reactions (3) and (4). They are comparable to reaction (1) on PG. Reaction (2) has a similar barrier as the corresponding reaction on PG.

		\subsubsection{N-defected graphene}
		Finally, reaction steps of methane decomposition (1)-(4) were simulated in a cell with a single in-plane carbon atom substituted with nitrogen. In this case, the free energy profiles (see \Cref{figure:pmf_diss}(c)) have a similar shape to one another. With the exception of CH$_2$, the profiles flatten out at RC between 2.28~\si{\angstrom} and 2.51~\si{\angstrom} after a transition state is reached.

		The energy barriers together with the characteristic value of the reaction coordinates for the dehydrogenation reactions described in this section are summarized in \Cref{table:diss}. For comparison, the literature values for the decomposition barriers of isolated gaseous species and the barriers for dehydrogenation of methane fragments on (111) surface of nickel are also given there. It is apparent that both nickel surface and graphene monolayer act as catalysts for the decomposition of methane, with nickel surface being the stronger catalyst. These findings shed light on the mechanisms of the growth process of the first graphene layer on nickel surface and then the growth of the subsequent second graphene layer in a CVD growth process involving the methane precursor.

	\subsection{Adsorption of CH$_n$ decomposition products}
	\label{section:adsorption}
	\begin{figure}
	\centering
	\includegraphics[width=3.37in]{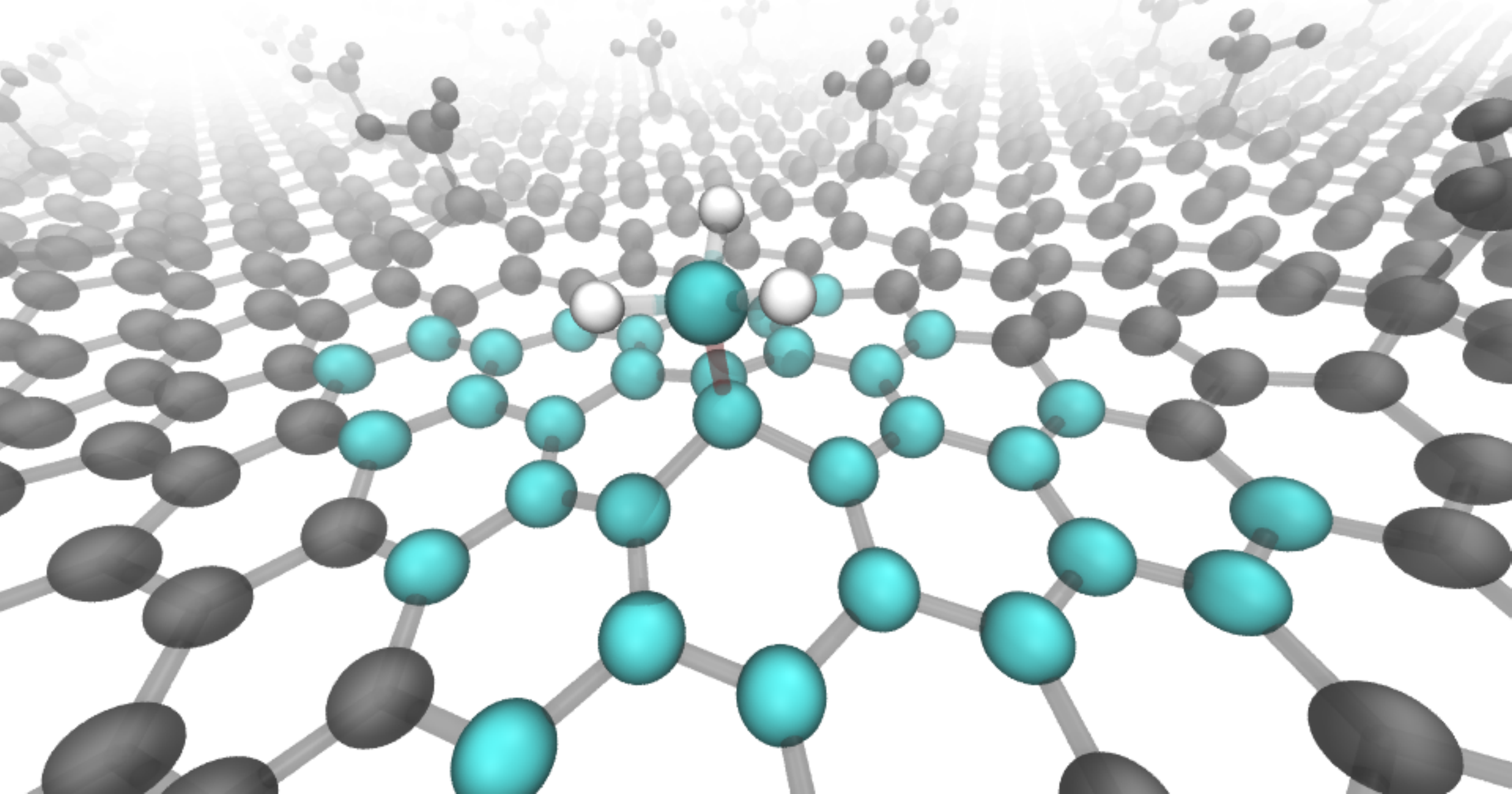}
	\caption{Equilibrium configuration of CH$_3$ on graphene. The adsorption RC is highlighted in red. The free energy profile for this reaction is plotted in \Cref{figure:PMF_ads} and labeled `CH$_3$'. The magnitudes of energy barriers and corresponding RC values are listed in \Cref{table:ads}. Color description: Colored atoms are included in the supercell (teal~--~carbon, white~--~hydrogen), and surrounding gray atoms are periodic images of the supercell. See Supplemental Information for graphics depicting configurations of other reaction steps on pristine graphene.}
	\label{figure:ch3_ads}
	\end{figure}

	\begin{table}
		\caption{Free energy barriers and reaction coordinates (RCs) for adsorption of methane gas constituents on pristine graphene (PG), SW-defected graphene (SWG) and nitrogen-substituted graphene (NSG) as calculated in present studies.}
		\centering
		\begin{tabularx}{\linewidth}{l|YYYYYYY}
			\hline
			Species & RC (\si{\angstrom}) of chemisorbed molecule & Desorption barrier (\si{\electronvolt}) & RC (\si{\angstrom}) of transition state & Activation energy for chemisorption (\si{\electronvolt}) \\
			\hline
			$CH_4@PG$  & --   & 0.00 & --    & 2.51   \\
			$CH_4@SWG$ & 1.60 & 0.16 & 1.81  & 1.84   \\
			$CH_4@NSG$ & --   & 0.00 & --    & 2.41   \\
			$CH_3@PG$  & 1.64 & 0.18 & 2.19  & 0.21   \\
			$CH_3@SWG$ & 1.60 & 0.42 & 2.31  & 0.05   \\
			$CH_3@NSG$ & 1.58 & 0.13 & 2.13  & 0.01   \\
			$CH_2@PG$  & 1.57 & 0.41 & 2.18  & --     \\
			$CH_2@SWG$ & 1.53 & 0.84 & 2.57  & 0.00   \\
			$CH_2@NSG$ & 1.56 & 0.11 & 2.31  & 0.00   \\
			$CH@PG$    & 1.44 & 0.55 & 3.14  & 0.10   \\
			$CH@SWG$   & 1.48 & 1.79 & --    & --     \\
			$CH@NSG$   & 1.45 & 0.35 & 2.33  & 0.05   \\
			$C@PG$     & 1.51 & 0.23 & 1.95  & 0.24   \\
			$C@SWG$    & 1.52 & 0.35 & 2.11  & 0.04   \\
			$C@NSG$    & 1.45 & 0.24 & 2.10  & 0.30   \\
			$H@PG$     & 1.14 & 1.00 & 2.84  & 0.01   \\
			$H@SWG$    & 1.12 & 0.96 & 2.00  & 0.02   \\
			$H@NSG$    & 1.10 & 0.85 & 2.80  & 0.16   \\
			\hline
		\end{tabularx}
		\label{table:ads}
	\end{table}
	In the previous section we have described the energetics of the chemical reactions leading to sequential decomposition of methane molecules. Here we focus on adsorption processes of CH$_4$ and fragments emerging from the previous dehydrogenation reactions, i.e., CH$_3$, CH$_2$, CH, C, and H, on pristine and defected graphene with Stone-Wales defects and substitutional nitrogen. To investigate the adsorption processes, we perform a series of simulations employing the computational scheme from \Cref{section:Methods_description}. In these simulations, we set the reaction coordinate as the distance between an in-plane C atom in graphene and the C or H atom in the adsorbed species (see \Cref{figure:ch3_ads} illustrating the geometry of CH$_3$ adsorbed at pristine graphene).

	Depending on the species, we observe direct chemisorption (or lack thereof in case of CH$_4$) either on the ``top'' or ``bridge'' site in graphene. Adsorption on ``top'' involves the creation of a single bond between a surface atom and the adsorbate. This is the case for species that have one unpaired electron, i.e.~CH$_3$, C and H. This configuration can be seen in \Cref{figure:ch3_ads}. Species adsorbed on the ``bridge'' site~--~CH$_2$ and CH~--~ have at least two unpaired electrons. Upon adsorption, two bonds are created and the species is adsorbed between in-plane carbon atoms as seen in \Cref{figure:ch3_diss}. Besides the two figures referenced in this section, atomic configurations for other reactions can be viewed in online supplemental materials (Figure~S3 and S4).

	Generally, the shape of free energy profiles is more complex compared to CH$_n$ decomposition steps. This is due to different interactions that determine bonding between the molecules and the surface. Moving from low to high RC values, the first and usually deepest minimum corresponds to chemisorption equilibrium. Here, covalent bonding is the strongest and dominates over other interactions. For higher values of RC, chemisorption becomes much weaker and non-covalent bonding via van der Waals forces leads to physisorption of some species.

		\subsubsection{Adsorption on ideal graphene}
		\label{section:ads_pristine}
		\begin{figure}
		\centering
		\includegraphics[width=3.37in]{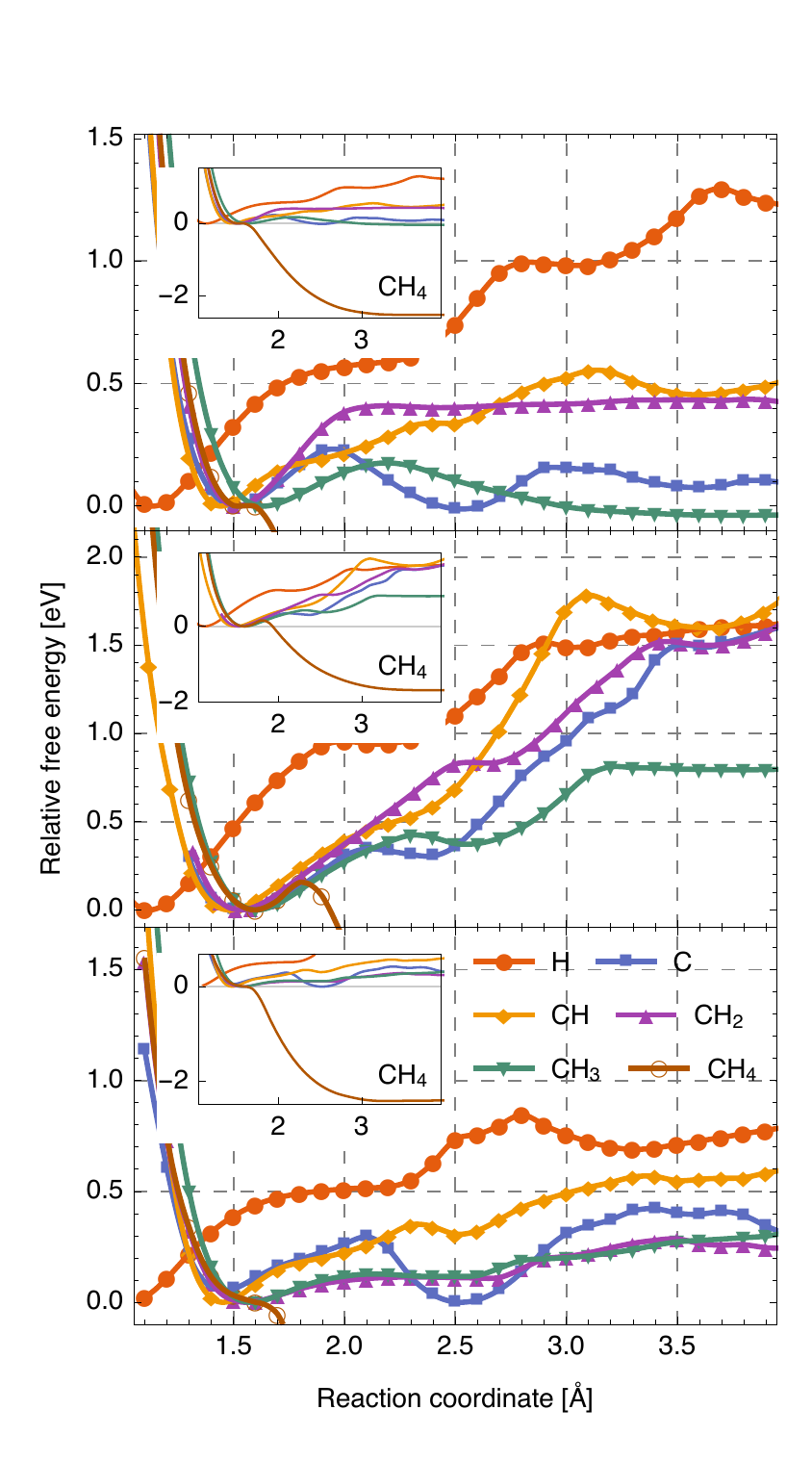}
		\caption{Relative free energy profiles of CH$_n$ and H adsorption on (a) pristine graphene, (b) graphene with a SW defect, (c) graphene with a N substitution. Insets depict the shape of CH$_4$ adsorption profiles which reach much lower values than the other profiles and are omitted from the main plots. Curves have been shifted so that zero energy corresponds to chemisorption (covalent bond) minima. The reaction coordinate is selected as the bond length between one of the in-plane carbon atoms and the carbon atom that is the center of a CH$_n$ molecule, as depicted in \Cref{figure:ch3_ads} for CH$_3$ adsorption reaction.}
		\label{figure:PMF_ads}
		\end{figure}

		We start the discussion of the adsorption processes considering adsorption of CH$_n$ ($n = 0-4$) and H to pristine graphene. The free energy profiles as the function of the reaction coordinate are depicted in \Cref{figure:PMF_ads}(a) and the energies of barriers and critical reaction coordinates are presented in \Cref{table:ads}. The listed barriers correspond to the free energy difference between the chemisorbed state and the transition state (desorption barrier) and between the physisorbed state and the transition state (activation energy for physisorption). The picture of adsorption processes that emerges from these calculations can be summarized as follows.

		$\mathbf{CH_4}$: The full methane molecule does not undergo stable chemisorption. Free energy declines monotonically with increasing RC. The lack of free electrons causes strong repulsion between the surface and the molecule. The energy barrier for CH$_4$ approach to the surface is estimated at around 2.51~\si{\electronvolt}.

		$\mathbf{CH_3}$: The methyl radical (CH$_3$) shows stable chemisorption. The barrier for desorption is calculated to be equal to 0.18~\si{\electronvolt}. There is another minimum on the other side of the transition state. It corresponds to a physisorbed state and its depth is similar to the chemisorption minimum -- the barrier in the other direction is equal to 0.21~\si{\electronvolt}. This shows that for this radical there is no strong dominance of one type of bonding.

		$\mathbf{CH_2}$: The barrier for desorption from a stable chemisorbed state is equal to 0.41~\si{\electronvolt}. This is roughly twice the barrier for CH$_3$ desorption. This is possibly due to the fact that CH$_2$ forms two covalent bonds with the surface (see \Cref{figure:ch3_diss}). We examine this effect in more detail in \Cref{section:density}. In this case, there is no stable physisorption, indicated by a flat energy profile above the transition state.

		\textbf{CH}: This radical with three unpaired electrons has a complicated adsorption free energy profile that cannot be interpreted as easily as in the case of CH$_4$, CH$_3$ and CH$_2$. The chemisorption minimum lies lower than for CH$_3$ and CH$_2$. Then the free energy rises to a transition state located at RC equal to 3.14~\si{\angstrom}. The barrier to reach the transition state from the chemisorption minimum is 0.55~\si{\electronvolt}. The slope of the profile is not constant and it flattens out at around 2.4~\si{\angstrom} with the relative free energy equal to ca. 0.3~\si{\electronvolt}. A second shallow minimum appears at RC=3.6~\si{\angstrom}. The height of the barrier for the reverse reaction is equal to 0.1~\si{\electronvolt}.

		\textbf{C}: For atomic carbon the chemisorption minimum is closer to CH$_2$ minimum. The first transition state is reached after climbing a barrier of 0.23~\si{\electronvolt}. This value is similar in magnitude to the CH$_3$ transition state but its RC is slightly lower at 1.95~\si{\angstrom}. C features two separated physisorption minima. One of them is similar in depth to the chemisorption minimum. The other, separated from the first by a barrier of 0.08~\si{\electronvolt}, is relatively broad and shallow.

		\textbf{H}: The chemisorption minimum is located at lower RC than for CH$_n$ molecules due to the shorter equilibrium C-H bond length. The shape of the free energy profile is similar to CH: a maximum that can be associated with a transition state lies at ca. 2.8~\si{\angstrom}. There is also a brief flattening at 2.12~\si{\angstrom} and 0.58~\si{\electronvolt}. There is a physisorption minimum above the transition state. The barrier to reach it from the chemisorbed state is equal to ca. 1~\si{\electronvolt}. There is another transition state at the free energy maximum at RC=3.7~\si{\angstrom}. The next minimum or plateau is not reached in the simulated RC range.

		Further, we have investigated how the adsorption reactions can be modified by the presence of defects, and we turn first to the adsorption of CH$_n$ molecules and radicals in the neighborhood of SW defects in graphene.

		\subsubsection{Adsorption on Stone-Wales-defected graphene}
		In general, adsorption free energy profiles on SW-defected graphene (\Cref{figure:PMF_ads}(b)) are more steep than on ideal graphene. This supports a previous finding that this defect induces stronger binding of methane decomposition products \cite{APPA}. Four of the profiles~--~for C, CH, CH$_2$ and CH$_3$ follow a similar shape after leaving the equilibrium state. They begin to diverge as different species reach their first transition state. However, they still follow each other closely. With the exception of CH$_3$ and CH$_4$, all species end up at similar relative free energy values at the end of the simulated RC ranges. For pristine graphene, the profiles look more divergent at high RC. However, the difference between the highest (H) and lowest (CH$_3$) relative energy (excluding CH$_4$) is similar for pristine and SW graphene.

		Let us now look how the free energy curves for the adsorption process change on nitrogen substitutional impurity.

		\subsubsection{Adsorption on N-defected graphene}
		Free energy profiles of CH$_n$ adsorption close to the nitrogen atom (\Cref{figure:PMF_ads}(c)) are either mostly unchanged (for C) or flattened (for H, CH, CH$_2$, CH$_3$) compared to pristine graphene. For C the difference lies mostly in the depth of the second minimum at 2.53~\si{\angstrom}. The barrier for desorption (in the direction of low to high RC) is higher than in case of chemisorption. CH$_3$ and CH$_2$ fragments are chemisorbed very weakly to the surface~--~their desorption is associated with the lowest desorption barriers among the studied surface geometries. These results suggest that the N substitution is associated with weaker covalent bonding at the defect, even though intuitively this region should be electron-rich and thus promote chemisorption. The profiles as a whole are significantly more convergent at high RC limit.

		To shed light on the mechanisms of the adsorption process, we have also investigated the changes of the electronic densities around the adsorbed species.

		\subsection{Electron density near adsorption sites}
		\label{section:density}
		\begin{figure*}
		\centering
		\frame{\includegraphics[width=.49\linewidth]{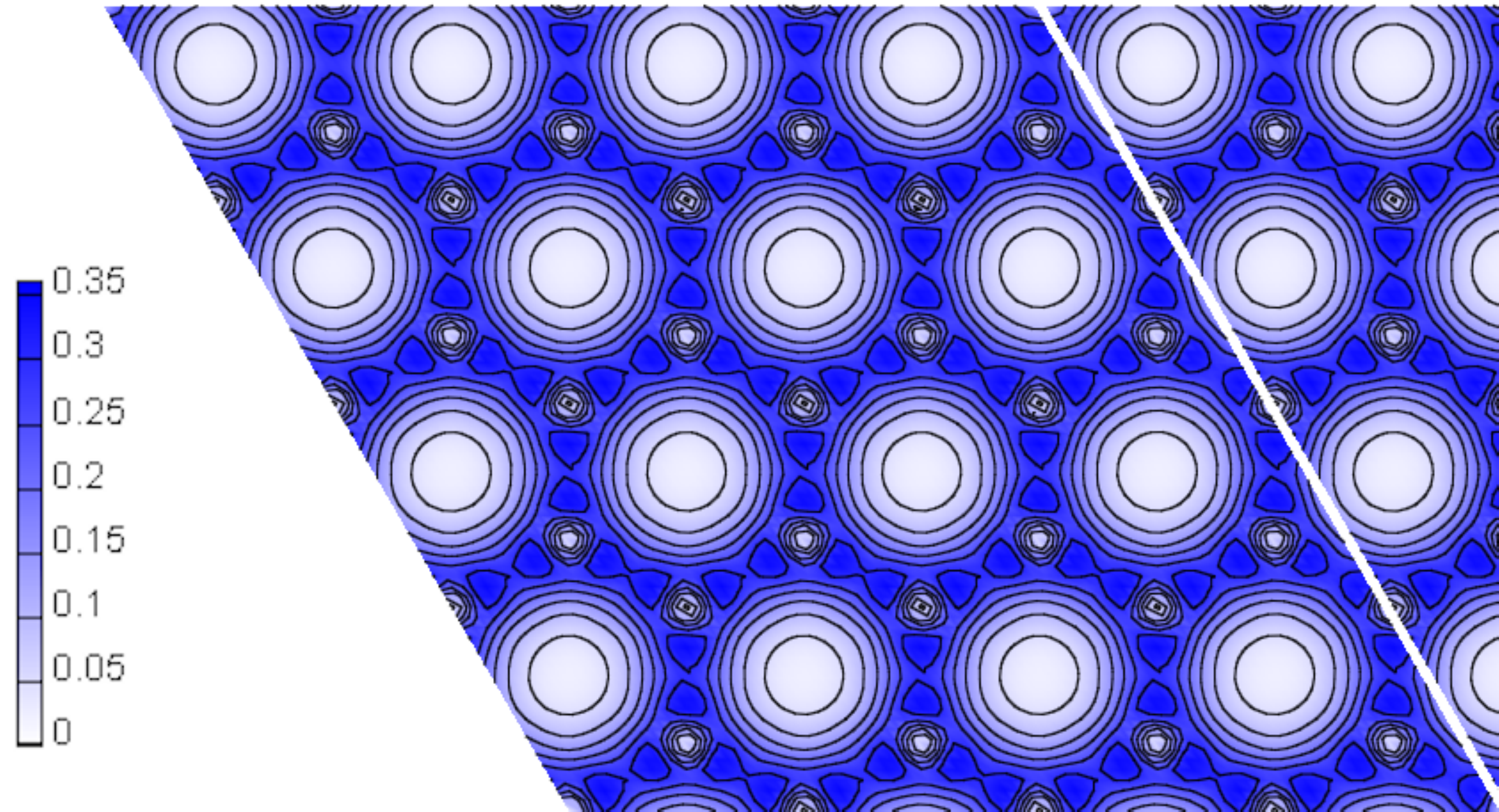}}\\
		\frame{\includegraphics[width=.49\linewidth]{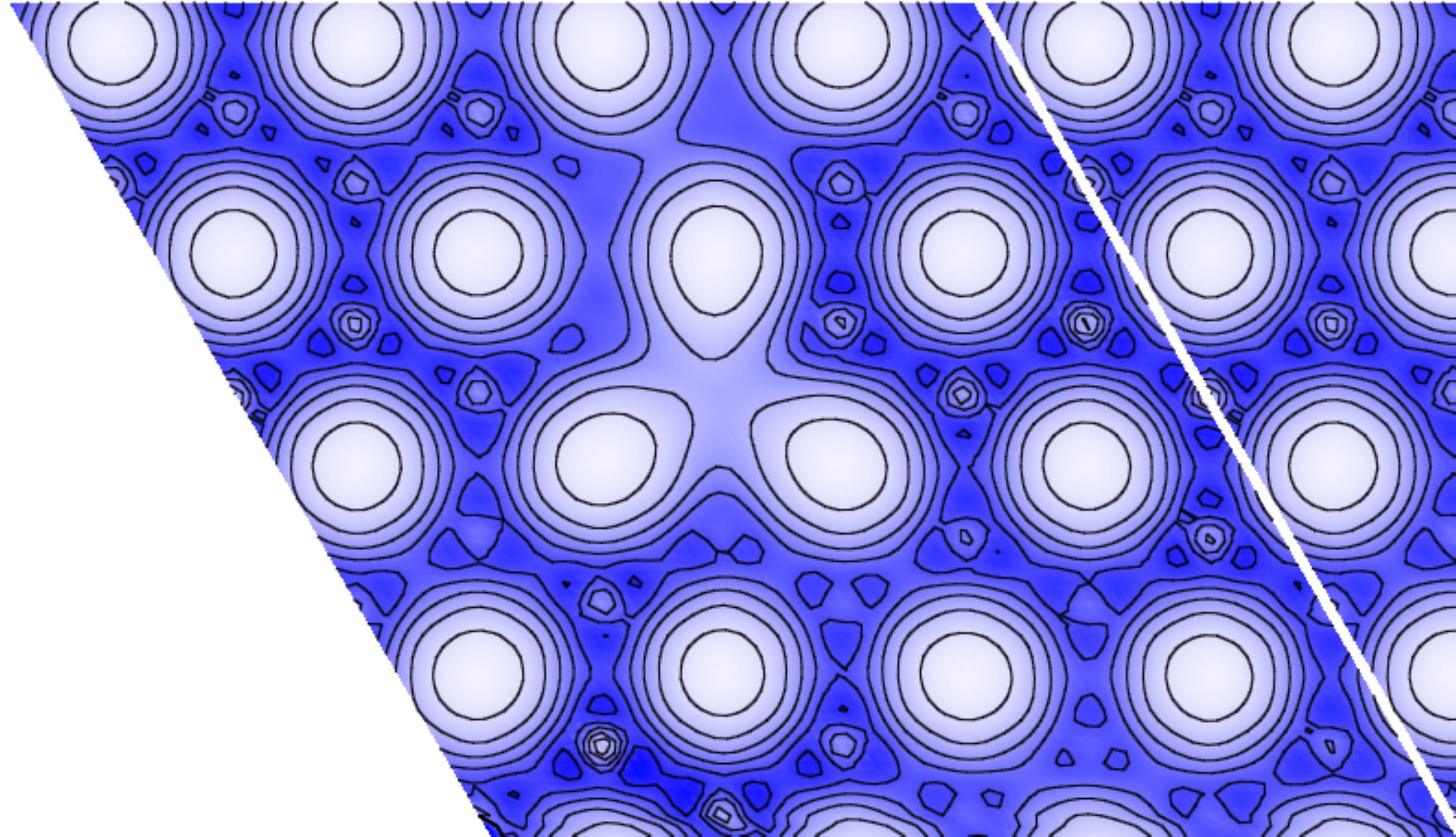}}
		\frame{\includegraphics[width=.49\linewidth]{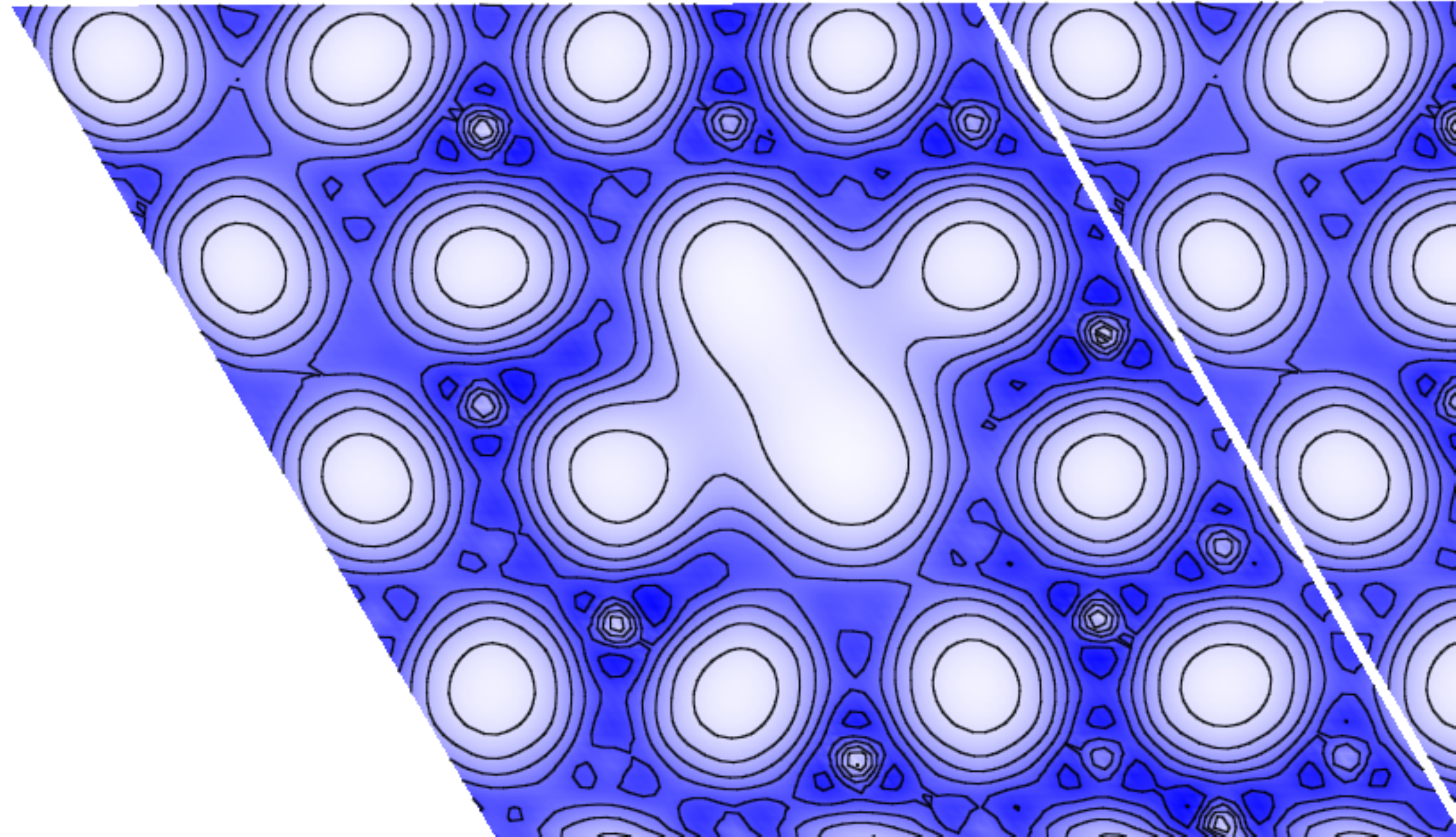}}
		\caption{Plots of the in-plane cross section of electron density (in $e/\si{\angstrom}^3$) in pristine graphene (top), graphene with a CH$_3$ admolecule (bottom left), and graphene with a CH$_2$ admolecule (bottom right).}
		\label{figure:rho}
		\end{figure*}
		
		In this section we examine how adsorption of different species influences electron density in the vicinity of adsorption sites. For pristine graphene, calculated density distribution forms an uniform hexagonal lattice that mirrors the atomic configuration of carbon atoms and the $\sigma$ bonds between them (see \Cref{figure:rho}). The regions in the middle of carbon rings are depleted of electrons and separated by electron-rich bonds.

		Upon chemisorption, some density is transferred from the surface to the adsorbed molecule. This creates an in-plane region of charge depletion. The effect is highly dependent on the type of admolecule. Two of the configurations, $CH_3@PG$ and $CH_2@PG$, analyzed in \Cref{section:ads_pristine}, are plotted in \Cref{figure:rho}. For CH$_3$, the adsorption site is a single in-plane carbon atom that forms a covalent bond with the carbon atom in the CH$_3$ radical. Density distribution is affected significantly only around the adsorption site.

		In the case of CH$_2$, the adsorption site is formed by two carbon atoms that create bonds via two unpaired electrons of the CH$_2$ radical. A large area of low electron density is formed through merging of electron-depleted regions in the middle of four hexagonal rings, just the ones that include the adsorption site atoms. Density distribution in the neighboring rings is affected to a greater degree than for CH$_3$. The change in electron density distribution is more pronounced than in the former case. This means that CH$_2$ adsorption and desorption processes induce a more dramatic change in the electronic structure of the graphene substrate. A similar conclusion can be drawn from data presented in \Cref{table:ads} and figures in \Cref{section:adsorption}. In case of pristine graphene and SW-defected graphene, the magnitudes of the energy barriers for CH$_2$ desorption are roughly two times higher than for CH$_3$. This also indicates that the CH$_2$ adsorption/desorption reactions are more complex. On the other hand, the same process on nitrogen-substituted graphene has a similar barrier for CH$_3$ and CH$_2$, which can be attributed to the fact that the N substitution itself influences the density distribution.

\section{Conclusions}
First principles molecular dynamics simulations with constraints have allowed us to determine the free energy landscape of dissociative adsorption of methane on graphene. We find that the presence of graphene, either pristine or with two types of defects, lowers the energy barrier for CH$_4$ decomposition in subsequent dehydrogenation steps compared to experimentally determined values for gaseous species. This indicates catalytic properties of graphene for the reaction of methane decomposition. Generally, the presence of defect does not change qualitative picture of decomposition and adsorption processes. In particular, we observe considerably low energy barrier for $CH_4 \rightarrow CH_3+H$ dehydrogenation reaction at Stone-Wales defect, similar in barrier reported for such reaction in the case of nickel surface used as catalyst. However, the energy barriers for all four steps of the $CH_4$ decompositions on Ni surface are much lower than in the case of graphene (pristine or defected), indicating that $CH_4$ the decomposition process is faster on Ni than on graphene. This explains the observed in experiments uniform growth of monolayer graphene on Ni surface (Frank-van der Merve like growth mode) and not growth of multilayered graphene islands (in the Volmer-Weber mode).

Free energy curves for adsorption exhibit more features than the ones for dehydrogenation. Due to the fact that interactions along the reaction paths are weaker, magnitudes of energy barriers are much lower. In this regime, van der Waals interactions play a significant role in the energetics of binding. Because of low energy barriers, adsorption of CH$_n$ species (n=3,2,1,0) on graphene appears to be volatile compared to dehydrogenation. In the dissociative adsorption process as a whole, the decomposition of methane gas most likely is the rate-limiting step.

\section*{Supplemental Information}
The online Supplemental Information for this paper provides justification of methodology with respect to Brillouin zone sampling and comparison of AIMD free energy profiles to profiles obtained from geometry optimization. Additional figures with atomic configurations of reaction products are also provided.

\section*{Acknowledgments}
The research has been performed in the framework of the ShaleSeq project (Physicochemical effects of CO2 sequestration in the Pomeranian gas bearing shales) funded / co-funded from Norway Grants in the Polish-Norwegian Research Programme operated by the National Centre for Research and Development, grant no.~POL-NOR/234198/100/2014. Calculations were performed at the ICM high performance computing center, University of Warsaw, grant no.~G60-8.

\bibliography{mwlazlo_jcp}
\end{document}